# Ultra-compact, high-NA achromatic multilevel diffractive lens via metaheuristic approach

**BUMIN K. YILDIRIM,**[1, *] **HAMZA KURT,**[2,3] **AND MIRBEK TURDUEV**[1]

[1]*Department of Electrical and Electronics Engineering, TED University, Ankara 06420, Turkey*
[2]*Department of Electrical and Electronics Engineering, TOBB University of Economics and Technology, Ankara 06560, Turkey*
[3]*School of Electrical Engineering, Korea Advanced Institute of Science and Technology (KAIST), Daejeon, 34141, Republic of Korea*
*\* bkagan.yildirim@tedu.edu.tr*

**Abstract:** Recently, the multilevel diffractive lenses (MDLs) have attracted considerable attention mainly due to their superior wave focusing performance; however, efforts to correct the chromatic aberration are still in progress. Here, we demonstrate the numerical design and experimental demonstration of high-numerical aperture (NA) (~0.99), diffraction-limited achromatic multilevel diffractive lens (AMDL) operating in microwave range 10 GHz-14 GHz. A multi-objective differential evolution (MO-DE) algorithm is incorporated with the three-dimensional finite-difference time-domain (3D FDTD) method to optimize both the heights and widths of each concentric ring (zone) of the AMDL structure. In this study, the desired focal distance ($\Delta F_d$) is treated as an optimization parameter in addition to the structural parameters of the zones for the first time. In other words, MO-DE diminishes the necessity of predetermined focal distance and center wavelength by also providing an alternative method for phase profile tailoring. The proposed AMDL can be considered as an ultra-compact, the radius is $3.7\lambda_c$ where $\lambda_c$ is the center wavelength (i.e., 12 GHz frequency), and flat lens which has a thickness of $\sim \lambda_c$. The numerically calculated full-width at half-maximum (FWHM) values are below $0.554\lambda$ and focusing efficiency values are varying between 28% and 45.5%. To experimentally demonstrate the functionality of the optimized lens, the AMDL composing of polylactic acid material (PLA) polymer is fabricated via 3D-printing technology. The numerical and experimental results are compared, discussed in detail and a good agreement between them is observed. Moreover, the verified AMDL in microwave regime is scaled down to the visible wavelengths to observe achromatic and diffraction-limited focusing behavior between 380 nm – 620 nm wavelengths.

## 1. Introduction

One of the first manipulation of light in the human history can be considered as the focusing/gathering of light into the desired location in the space. In this regard, starting from ancient times the studies on the focusing of light were fundamental for the light-matter interaction phenomenon [1]. To achieve light focusing, different types of curved structures called as lenses are studied. Since the first designed lenses, the optical aberrations such as monochromatic and chromatic, are still considered as one of the main hurdles of the focusing of light [2]. If the light having different wavelengths focused into different points this effect is called chromatic aberration. This effect appears due to material refractive index dependence on wavelength of the optical device (simple lenses, diffractive lenses, multilevel-diffractive lenses) [3, 4]. The creative endeavors to reduce the chromatic dispersion started with achromatic doublets in the 18[th] century [5], followed by applying hybrid refractive-diffractive lenses [6, 7] and graded-index (GRIN) optics [8].

Moreover, contemporary studies also brought different solutions to overcome chromatic dispersion. One of the most common methods is to design metalenses by tailoring the phase profiles [9-13], combining the metasurfaces as a hybrid structure [14], manipulating the polarization rotation of incident light [15], and metasurfaces doublets [16]. Additionally, multilevel diffractive lenses (MDLs) provide not only achromatic focusing but also polarization-insensitive, broadband, and energy efficient focusing with their reduced weight and size compared to the metalenses [17-21]. In the MDL structures, each zone plays an important role on chromatic aberration-free focusing effect if the zone's parameters are adjusted properly [22]. Hence, the various optimization and inverse design methods have been exploited to obtain appropriate MDL parameters for efficient focusing [23-26]. One of the important design characteristics of the metalenses and MDLs is their numerical aperture (NA) value which is a function of focal length, the radius of the lens, and the refractive index of the medium. It is difficult to obtain both high NA value and chromatic aberration-free focusing [27]; however, an appropriate design of MDL can overcome low efficiency and chromatic dispersion problems of the metalenses [28, 29].

In this paper, we present high-NA, diffraction-limited broadband achromatic multilevel diffractive lens (AMDL) operating in the microwave regime between 10 GHz - 14 GHz frequencies. A multi-objective differential evolution (MO-DE) algorithm is applied to minimize chromatic aberration and maximize the focusing efficiency of the AMDL by tightly confining the diffracted light energy in the main lobe and suppressing the side lobes (suppressing side lobes are aimed to achieve clear and efficient focal spot). Here, the heights and the widths of each concentric ring (i.e., zones) of the AMDL are optimized to achieve a predefined objective function. Moreover, differently from the proposed achromatic metalens and MDL designs present in the literature, in this study the focal distance is directly optimized by the algorithm without fixing it to the desired center frequency (fixing of the center frequency is an important requirement for phase profile engineering in the achromatic lens designs). In other words, this approach diminishes the necessity of determining focal distance and center wavelength and provides an alternative to engineer phase profiles. In addition, to experimentally verify the numerical results, measurements in the microwave regime are performed by fabricating proposed AMDL using 3D - fused deposition modelling (3D-FDM) with polylactic acid (PLA) material. We also scale down the optimized AMDL to the visible wavelengths by taking 550 nm as a reference wavelength ($\lambda_{ref}$) to observe the achromatic focusing behavior. All AMDL characteristics in wavelengths between 380 nm – 620 nm are investigated in detail in the discussion section.

## 2. Design Methodology and Numerical Results of AMDL

In this study, the Differential Evolution (DE) algorithm is used to obtain an ultra-compact, high- NA and broadband functional achromatic MDL. Here, DE algorithm is chosen as a design method due to its previous achievements in the earlier studies [4,30-32]. The DE algorithm can be considered as an evolutionary algorithm which mimics the natural evolution and selection processes for survival of the fittest function. This algorithm iteratively searches for candidate solutions for a specific design problem to find the optimum solution by utilizing its bio-inspired mechanisms like cross-over, mutation, and selection [33]. In the beginning of the optimization process, the randomly generated individuals which are called as parents are started to examine and they are updated in an iterative manner to minimize the predefined cost function. During the optimization process, next generations (i.e., children) will be created from the best candidate among the parents via bio-inspired mechanisms. This process continues until the algorithm reaches the local minima of the cost function.

The multi-objective DE (MO-DE) optimization algorithm is embedded into 3D finite-difference time domain (FDTD) [34] method to design ultra-compact, high-NA and achromatic MDL structure. Fig. 1(a) illustrates the AMDL structure intended to design with its structural parameters (i.e., heights (*h*) and widths (*w*)) and achromatic focusing strategy where the colored parallel lines refer to polychromatic light excitation. Here, MO-DE optimizes these parameters (i.e., *h* and *w* are determined by DE algorithm) to reveal both achromatic (focusing of polychromatic light into a particular point) and diffraction limited focusing (forcing both the full-width at half-maximum (FWHM) and maximum side-lobe level (MSLL) values to decrease) by using user defined multi-objective cost function. The analytical expression of the defined multi-objective cost function is given in Eq. (1):

$$f_{cost} = w_1 \times \sum_{i=1}^{n} FWHM(\lambda_i) + w_2 \times \sum_{i=1}^{n} MSLL(\lambda_i)$$
$$+ w_3 \times \sum_{i=1}^{n} |\Delta F(\lambda_i) - \Delta F_d| + w_4 \times \sum_{i=1}^{n/2} |\Delta F(\lambda_{n-1+i}) - \Delta F(\lambda_i)|, \quad (1)$$

Here $w_1$, $w_2$, $w_3$, and $w_4$ values are defined as weighting factors to adjust the balance between each goal in the function. It is evident that the higher the value of a weighting factor is, the more the significance of the goals increases. However, the exact values of the weighting factors should be assigned before the optimization process. It is well known that the design of any optical device by using optimization algorithms requires an enormous amount of computational time due to initially unknown parameters of the optimization method (in our case weighting factors). In In this study, to determine the appropriate objective function with appropriate weighting factors, a great number of trial simulations were performed. In this regard, by analyzing the data obtained from the trial simulations, we defined the objective function with consistent weighting values. The first and second terms of the function are related with the focusing performance of the AMDL and the remaining terms are responsible for correcting the focusing chromaticity issue by equalizing all back focal distances of the different wavelengths. In Eq. (1), $FWHM(\lambda_i)$ and $MSLL(\lambda_i)$ are the FWHM and MSLL values of each wavelength ($\lambda_i$) corresponding to microwave frequency regime 10 GHz – 14 GHz. Within this frequency interval 15 equally distant frequency (i.e., 15 different wavelengths) points are chosen for $f_{cost}$ calculation. $\Delta F(\lambda_i)$ represents the back focal distance of the $i^{th}$ wavelength and $\Delta F_d$ is the desired focal length. Here, desired focal distance $\Delta F_d$ is determined by MO-DE algorithm during the optimization process. It is important to note that, the extremum values of $\Delta F_d$ that can be taken by the MO-DE algorithm lies between $0.2\lambda_c - 1\lambda_c$ (this interval defined by considering near field focusing and to achieve high-NA) where $\lambda_c$ is the center wavelength (i.e., 12 GHz frequency). This is very important difference from the other achromatic lens design strategies in the literature [12-20, 25-29] since here optimization method decides where the better focal point location is beneficial for achieving achromatic focusing effect. In short, the aim of the optimization algorithm is to minimize $f_{cost}$ function by reducing FWHM and MSLL values for all frequency range and focusing all wavelengths into the same focal point again decided by the MO-DE.

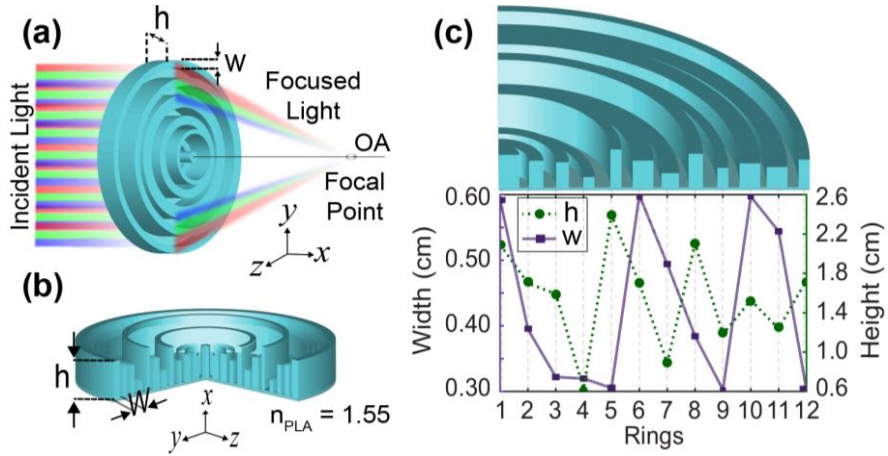

Fig. 1. (a) The schematic representation of planned AMDL with light focusing behavior and optimization parameters. (b) Perspective view of designed AMDL formed of PLA and (c) the quarter-cross sectional view of the lens with the plot of height and width of each ring.

The designed lens is composed of PLA material which has low permittivity value $\varepsilon_{PLA} = 2.4025$ in the microwave frequencies [35]. The number of zones (concentric rings) whose structural dimensions vary is fixed to 12. This number is intentionally chosen according with fabrication limitation of the 3D printing physical area which is limited to 29 cm ×19 cm ×16 cm (physical area of the printing of "MakerBot Replicator +" 3D printer). Moreover, the range of values of the zones' heights and widths that can be defined by the MO-DE are changing from 3 mm to 6 mm and from 6 mm to 26 mm, respectively. Finally, after exact defining of multi-objective function with all parameters, the MO-DE is integrated with 3D FDTD and . The finally optimized AMDL structure is presented in Fig. 1(b) and the optimization process is nearly 200 iterations (approximately 14 days). Here, in each iteration, 20 individual simulations are performed (20 different populations), i.e., 20 different AMDL structures are created. Moreover, during the optimization of the structural parameters, the numerical modelling and time-domain analysis of the AMDL structure in each iteration were carried out by the 3D FDTD method. In all numerical simulations, the computational domain was restricted to a 3D spatial domain and elimination of undesired wave back reflections are eliminated by perfectly matched layers [36]. The structure under optimization is excited with a broadband transverse-magnetic (TM- the magnetic field is in the *xy*-plane ($H_x$, $H_y$) and the electric field $E_z$ is perpendicular to the *xy*-plane;) polarized continuous sources with a Gaussian profile to calculate $f_{cost}$.

The MO-DE minimizes $f_{cost}$ function to achieve high-NA and achromatic MDL structure by optimizing the heights and widths of each zone. These parameters of each zone are presented in Figs. 1(b) and 1(c), respectively. The ultimate radius and thickness values of the AMDL are emerged as 9.3 cm and 2.6 cm where the substrate thickness is fixed to 0.2 cm, respectively. The substrate is required to hold optimized zones to generate AMDL structure. As can be seen from Fig. 1(c), the final values of zone widths vary between 0.30 cm and 0.59 cm, while the corresponding heights changes from 0.63 *cm* to 2.40 *cm*. The MO-DE algorithm determined the desired focal length ($\Delta F_d$) as 13.82 mm ($0.553\lambda_c$ where $\lambda_c = 24.98 mm$ corresponds to 12 GHz frequency). As a result, optimization attempted to focus all wavelengths into the desired focal distance of 13.82 *mm*. Moreover, it also tried to increase the focusing performance of AMDL for all wavelengths (between 10 GHz–14 GHz interval) by preserving achromatic focusing effect.

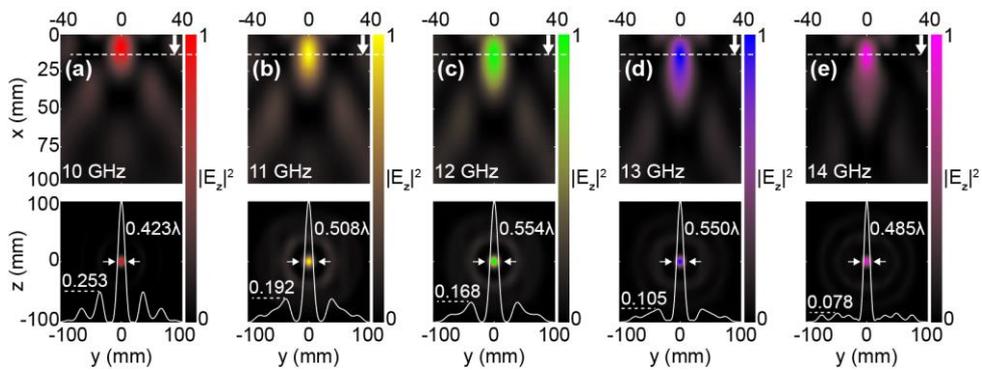

Fig. 2. The numerically calculated electric field intensity distributions (top) and calculated intensity distributions around the focal points with their lateral cross-sectional profiles and corresponding FWHM and MSLL values (bottom) at (a) 10 GHz, (b) 11 GHz, (c) 12 GHz, (d) 13 GHz, and (e) 14 GHz. The white horizontal dashed line indicates the desired focal distance ($\Delta F_d = 13.82 mm$) and the white arrows indicate the propagation direction.

Figure 2 shows the numerically calculated steady-state spatial electric field intensity ($|E_z|^2$) profiles of the focused beam in the *xy*- and *zy*-planes for different operating frequencies. In Figs. 2(a)-2(e) the electric field intensity distributions (depicted in top line) in *xy*-plane at the frequencies of 10 GHz -14 GHz with 1 GHz step are presented, respectively. Here, numerically calculated back focal distances are as follows 9.92 *mm*, 13.27 *mm* for 10 GHz and 11 GHz, respectively. On the other hand, AMDL structure focuses light at operating frequencies of 12 GHz, 13 GHz, and 14 GHz to the exact same back focal distance of 14.42 *mm.* In the same Figs. 2(a)-2(e), the figure plots in a bottom line illustrate the focal spot (field concentration) of calculated electric field intensity at the selected frequency points in *zy*-plane. The normalized lateral cross-sectional profiles of focal spot in *zy*-plane are also superimposed in these plots. As can be seen from Fig. 2(a)-2(e), the calculated FWHM values of focused beam vary between $0.423\lambda$ and $0.554\lambda$, where $\lambda$ denotes the wavelength of each selected frequency, while the corresponding MSLL values changes between 0.078 and 0.253. which means that most of the energy is confined in the focal spot (i.e., main lobe). If we look to the results presented in Fig. 2, we can see that the MO-DE achieved to focus the light near self-defined $\Delta F_d$ focusing distance of 13.82 mm. For instance, for operating frequencies of 10 GHz and 11 GHz the difference between self-defined (optimized) $\Delta F_d$ and obtained focal distances are $0.156\lambda_c$ and $0.022\lambda_c$, respectively. Also, for the rest of frequencies (12 GHz, 13 GHz, and 14 GHz) the shift from the $\Delta Fd$ is $0.022\lambda_c$ where $\lambda_c = 24.98 mm$. From the given results, we can see that implemented optimization method accomplished the strong focusing of light (FWHMs are near $0.5\lambda$) and achromatic focusing with the average shift in focal lengths of 2.2% for the frequencies of 11 GHz, 12 GHz, 13 GHz, and 14 GHz [25].

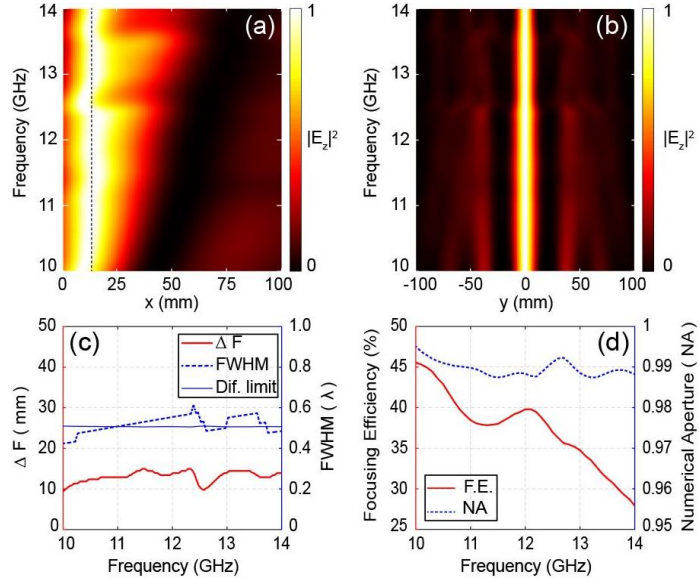

Fig. 3. The map of cross-sectional electric field intensity distributions in (a) longitudinal direction on the optical axis and in (b) lateral direction at focal points. The black vertical dashed line is representing desired focal distance ( $\Delta F_d = 13.82 mm$ ) (c) The plot of back focal distance ($\Delta F$) in mm (red-solid) and FWHM (blue-dashed). The blue solid line indicates the diffraction limited FWHM values. (d) The focusing efficiency (red-solid) and numerical aperture (blue-dashed) graphs of selected frequency interval.

To analyze broadband operation and performance of the designed AMDL, Fig. 3 is prepared. The map of cross-sectional electric field intensity ($|E_z|^2$) profiles in longitudinal direction along the lateral direction at the focal point location are given in Figs. 3(a) and 3(b), respectively. The dashed line superimposed to the map in Fig. 3(a) defines the location of $\Delta F_d$. Moreover, both maps in Figs. 3(a) and 3(b) demonstrate the strong focusing effect with corrected chromaticity through the broadband frequency regime except at the frequencies around 12.5 GHz where the focused light is slightly shifted from the dashed line.

There is a smooth transition from the general trend (being achromatic) to a wavelength dependent focusing at around 12.5 GHz. Both focal point and FWHM reach a minimum value at this frequency and the achromatic response is apparent as we continue tracing the rest of the bandwidth. It is surprising that the deployed algorithm generates an AMDL structure holding four zones and binary gratings of different heights and widths in a rather regular fashion. However, some of the structural parameters break the design rules of diffractive lens that can be used based on the analytical expressions [22]. Hence, even though it is difficult to identify the exact physical mechanism behind the characteristic observed at around 12.5 GHz, weakly guided resonance mode effect could be attributed to this behavior. When we calculated the transmission coefficient (the plot is skipped for simplicity) we observed a slight dip in the transmitted light (at exactly 12.68 GHz). It could be the indication of the light coupling to weakly guided mode in the AMDL structure. By inspecting the field plots in the xy plane, $(w/c) \times n_{eff} \times D$ becomes integer multiples of $2\pi$ at the resonance condition. The diameter of the AMDL is represented by $D$. Finally, the spatial periodicity, $\Lambda$ satisfies the condition $(\Lambda \ll \lambda)$ so that multi-level grating can be assumed as a subwavelength grating.

In the Fig. 3(c) calculated focal distances ($\Delta F$) and FWHM values are plotted for the frequencies between 10 GHz -14 GHz. The minimum and the maximum focal distance values are calculated as 9.92 *mm* and 15.02 *mm*, respectively and the variation is only 5.10 mm which almost equals to $\lambda_c/5$, where $\lambda_c$ is the wavelength of the center frequency (12 GHz), which is 24.98 mm. The average value of the broadband $\Delta F$ is equal to 13.49 mm which means that the average shift in focal lengths equals to 2.38% [25]. All the presented results verify that the implemented MO-DE algorithm satisfactorily suppresses the focal shift over a broadband range. On the other hand, the calculated FWHM values are varying between $0.423\lambda$ and $0.610\lambda$ for the frequency interval between 10 GHz and 14 GHz. We know that the theoretical diffraction limited FWHM of the focused light is $\sim \lambda/(2 \times NA)$, and we can see that in both smaller and larger frequency values designed AMDL structure surpasses this limit. To further analyze the focusing performance of AMDL within the frequency region of 10 GHz -14 GHz numerical aperture (NA) and focusing efficiency values are calculated. A common approach for calculation of focusing efficiency, which is taking the proportion of the collected light on the focal plane (exactly on the focal spot location) with a radius defined as three times that of the FWHM of a spot size, is used [37, 38].

It is also important to analyze the broadband focusing performance of the proposed AMDL quantitatively. For this reason, the focusing efficiency and NA values are extracted and demonstrated in Fig. 3(d). Along the defined frequency interval, NA values are calculated as above the 0.986 and the focusing efficiency changes from 28% to 45.5%. As can be seen from the corresponding figure, the overall broadband focusing efficiency starts decreasing linearly after frequency of 12 GHz from 40% to 28%. If we look to the light focusing map presented in Fig. 3(a), one can see that the light is tightly confined around the focal spots until 12 GHz but after even though light is focused the overall energy is dispersed around the focal spot. Moreover, low focusing efficiency of the AMDL can be associated with a shadowing effect around the focal point which mainly occurs due to the finite depth surface [39, 40]. In other words, for the case AMDL structure the discontinuities of the wrapped phase due to different heights of the zones as well as the height difference between zones and substrate, leads to creation of a shadow that wastes light into undesired artificial orders [41, 42].

### 3. Experimental Analysis of Achromatic Focusing Effect

To verify the operation principle and focusing performance of the designed AMDL, experiments are conducted in the microwave frequency regime. The designed lens is produced by using 3D – FDM with polylactide (PLA) thermoplastic polyester material. Here PLA material is all dielectric, low-cost and biodegradable material which has a low dielectric constant value of $\varepsilon_{PLA} = 2.4025$ at microwave regime between 10 GHz-14 GHz. The designed AMDL structure is 3D printed by using commercially available "MakerBot Replicator +" 3D printer. Furthermore, FDM of the 3D printer is set to fabricate the AMDL structure with 100% infill ratio to generate the solid and homogeneous distribution of PLA material throughout the AMDL structure.

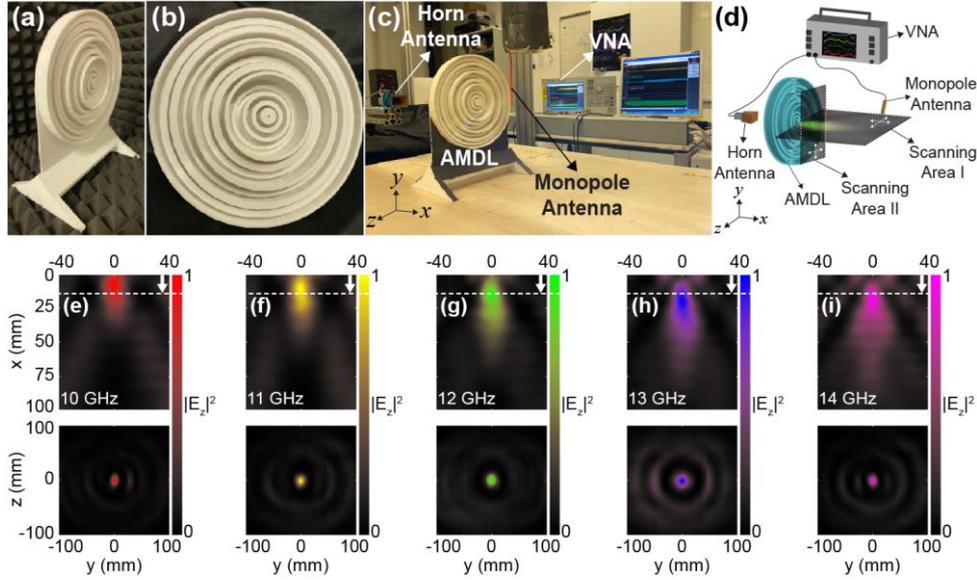

Fig. 4. The photograph of the 3D printed AMDL from (a) perspective and (b) top view. (c) The photographic view of the experimental setup. (d) The schematic representation of experimental setup to scan electric field intensity distributions at back focal plane (*xz*-plane – Scanning Area I) and around the focal point (*zy*-plane - Scanning Area II). The experimentally measured field intensity profiles (top) and measured intensity distributions around the focal points (bottom) at (e) 10 GHz, (f) 11 GHz, (g) 12 GHz, (h) 13 GHz, and (i) 14 GHz. The white horizontal dashed line indicates the desired focal distance ($\Delta F_d = 13.82\,mm$) and the white arrows indicate the propagation direction.

The photographic illustration of the perspective and top views of the fabricated AMDL lens are presented in Figs. 4(a) and 4(b), respectively. Also, we produced stand for the AMDL structure to keep it straight (in parallel to the horn antenna that illuminates AMDL with microwave source) as can be seen from perspective view of AMDL in the Fig. 4(a). Produced stand was covered by microwave absorbing materials to reduce the undesirable reflections. In the microwave experiments, Agilent E5071C ENA Vector Network Analyzer (VNA) is used to produce Gaussian profiled electromagnetic wave source through a horn antenna connected to VNA, whereas a monopole antenna is employed to scan the electric field distributions. The photographic illustration of the experimental setup and schematic view of the experimental measurements are presented in Figs. 4(c) and 4(d), respectively. In Fig. 4(d), schematic of electric field distribution scanning areas is presented and defined as Scanning Area I for *xz*-plane and Scanning Area II for *zy*-plane. The electric field distributions of the corresponding scanning areas are measured by using motorized stage which moves in *xz*- and *zy*- directions with the steps of 2 *mm*.

Experimentally measured electric field distributions at the focal plane (*xz*-plane) with corresponding focal spot distributions (*zy*-plane) are presented in Figs. 4(e)-4(i) for the selected frequencies of 10 Ghz, 11 GHZ, 12 GHz, 13 GHz, and 14 GHz. In the given Fig. 4, white dashed line defines optimized desired focal distance $\Delta F_d$ position. Corresponding focal distances for the frequencies of 10 GHz, 11 GHZ, 12 GHz, 13 GHz and 14 GHz are as follows 7.52 *mm*, 10.57 *mm*, 11.47 *mm*, 19.65 *mm*, and 16.96 *mm*, respectively. Also, measured FWHM values remain below the $0.8\lambda$ for [10 GHz, 11 GHz, 12 GHz, and 14 GHz] except the 13 GHz ($0.864\lambda$). The normalized MSLL values are changing between 0.145 and 0.3827. The measured numerical aperture values above the 0.975 for the frequency interval of 10 GHz – 14 GHz. As can be seen from the given results in Fig. 4, the numerically calculated and experimentally measured focal distance values are not the exact same but have the similar tendency. Dissimilarity can be caused by the utilized possible nonideal conditions of the experimental measurements: horn antenna does not generate the same Gaussian profiled electromagnetic waves used in FDTD simulations, for the impurity of the material, the production capability of the 3D printer, external factors such as temperature and imperfect alignment of the horn antennas, etc. The experimentally measured results verify overall functionality and efficient performance of the AMDL despite the small discrepancies in the focal distances. Still AMDL shows the high-NA and strong focusing effect.

To verify broadband operation functionality of the designed AMDL, the maps of the experimentally measured cross-sectional intensity profiles on the optical axis and on the lateral direction of focal point are presented in Figs. 5(a) and 5(b), respectively. The similarity between the numerical calculations and experimental measurements are evident except the small shifts in the $\Delta F$ values and little decrease in the FWHM values. The quantitative demonstration of $\Delta F$ and FWHM values can be seen in Fig. 5(c). The minimum and the maximum focal distance values are calculated as 7.50 mm and 20.9 mm, respectively and the variation is only 13.40 mm which almost equals to $\lambda_c/2$, where $\lambda_c$ is the wavelength of the center frequency (12 GHz), which is 24.98 mm. The measured average value of the broadband $\Delta F$ is equal to 12.85 mm which means that the average shift in focal distances equals to 7.01% [25]. As can be seen from Fig. 5(c), measured FWHM values are varying from $0.68\lambda$ to $0.89\lambda$ which still show strong focusing effect of the proposed AMDL.

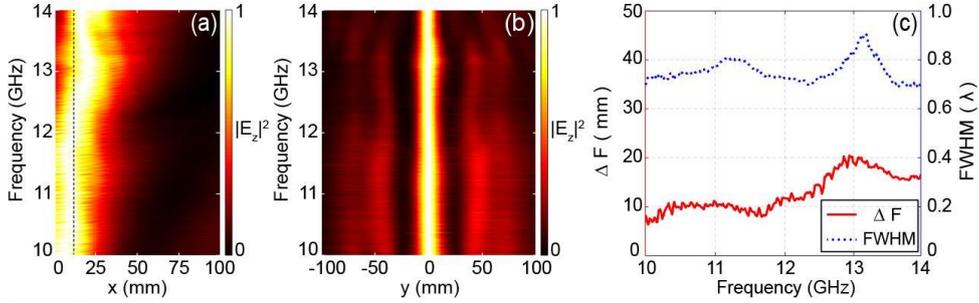

Fig. 5. The maps of experimentally measured cross-sectional intensity profiles (a) on the optical axis and (b) at focal points. The black vertical dashed line is representing desired focal distance ($\Delta F_d = 13.82\,mm$) (c) The graphs of both $\Delta F$ and FWHM values.

It is difficult to directly measure the focusing efficiency of the AMDL structure through our microwave experimental setup, thus we used an indirect method which resembles the numerical calculations [37, 38] to calculate the focusing efficiency of fabricated lens to provide comparable data. Here, the transmission value falling into a circular aperture with a radius equal to three times the FWHM centered at focal point is proportioned the transmission value falling into whole focal plane (200 mm × 200 mm). The obtained results are provided in Table 1. The experimentally measured focusing efficiency (F.E.) values are are almost 10% below the numerically obtained ones. The reason may be that the proportioned focal plane area is larger than the lens coverage (186 mm × 186 mm), unlike 3D FDTD simulations.

**Table 1. Numerical and Experimental Characteristics of AMDL with Their Average Values**

| | 3D FDTD | | | | | EXPERIMENT | | | | |
|---|---|---|---|---|---|---|---|---|---|---|
| Frequency (GHz) | $\Delta F$ (mm) | FWHM ($\lambda$) | MSLL ($\lambda$) | NA | F.E. (%) | $\Delta F$ (mm) | FWHM ($\lambda$) | MSLL ($\lambda$) | NA | F.E. (%) |
| 10 | 9.92 | 0.423 | 0.253 | 0.994 | 45.55 | 7.52 | 0.701 | 0.351 | 0.997 | 32.19 |
| 11 | 13.27 | 0.508 | 0.192 | 0.990 | 38.23 | 10.57 | 0.789 | 0.382 | 0.994 | 30.14 |
| 12 | 14.42 | 0.554 | 0.168 | 0.988 | 39.94 | 11.47 | 0.736 | 0.275 | 0.992 | 26.03 |
| 13 | 14.42 | 0.550 | 0.105 | 0.988 | 35.41 | 19.65 | 0.864 | 0.230 | 0.979 | 23.18 |
| 14 | 14.42 | 0.485 | 0.078 | 0.988 | 27.91 | 16.96 | 0.762 | 0.277 | 0.983 | 22.35 |
| Average | 13.29 | 0.504 | 0.159 | 0.990 | 37.41 | 13.23 | 0.770 | 0.303 | 0.989 | 26.78 |

To compare the results of the numerical calculation and experimental measurements, all data is collected in the Table 1. When we examine the data shared in Table 1, we can conclude that the experimental and numerical results are in good agreement with each other. Considering the average data (last column in Table 1), NA values are almost the same in both cases. The differences in other parameters are acceptable and it is possible to speculate that they are caused by problems such as exciting the structure with a partial plane wave due to finite aperture size of the horn antenna and the alignment problems of the antenna and the AMDL structure.

## 4. Discussion: Express Analysis of AMDL in Visible Wavelengths

It is also interesting to investigate the focusing performance of the optimized AMDL structure in the visible range. Unfortunately, we have no chance to experimentally fabricate and analyze the proposed AMDL in the visible wavelengths because of the absence of the experimental facility. Nevertheless, we scaled the structural parameters of the designed AMDL into visible wavelengths to analyze its performance in the visible regime.

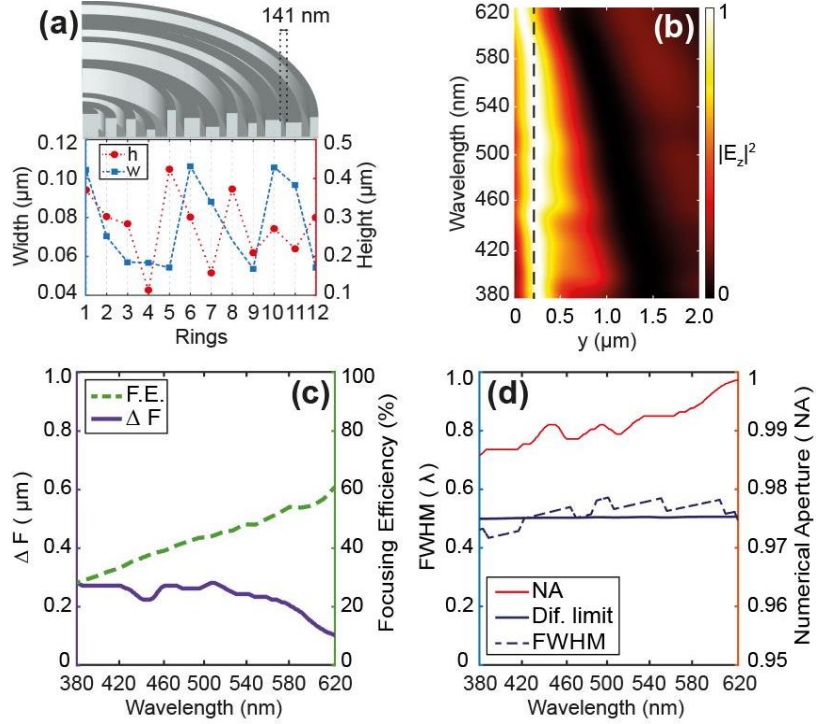

Fig. 6. (a) The quarter-cross sectional view of the nano-scaled lens with the plot of height and width of each concentric ring. The maps of cross-sectional intensity profiles on the optical axis, (b) plots of $\Delta F$ and focusing efficiency (c) and the graphs of NA and FWHM values between 380nm and 620nm.

For this reason, we chose 550 nm as a reference wavelength ($\lambda_{ref} = 550\,nm$) to scale down the AMDL and the refractive index of the lens material is fixed to 1.467 for visible wavelengths [43]. The heights and widths of each zones of AMDL are plotted in Fig. 6(a) with quarter cross-sectional view. In this case, the overall radius of the lens and the thickness of the lens become 1.652 μm ($3\lambda_{ref}$) and 0.46 μm ($<\lambda_{ref}$), respectively. The initial assigned distance between the closed edges of each zone is emerged as 141 nm as shown as an inset in Fig. 6(a). The achromatic focusing performance of the AMDL can be seen from cross-sectional intensity map in Fig. 6(b). Although, the maximum deviation of focal points is calculated as 180 nm ($\sim \lambda_{ref}/3$), the back focal distances remain almost constant ($\sim 270\,nm$) between wavelengths of 380 nm - 580 nm. In the range of wavelengths between 380 nm and 620 nm, the focusing efficiency values are linearly increasing from 28% to 60.7%. The calculated focal distance ($\Delta F$) and focusing efficiency values are superimposed in Fig. 6 (c), whereas the NA values and FWHM values in the same wavelength interval are shown in Fig. 6 (d). As expected, NA values are calculated above the 0.985 and FWHM values are almost equal to the corresponding diffraction-limited values. The FWHM values are varying between 0.464$\lambda$ and 0.564$\lambda$.

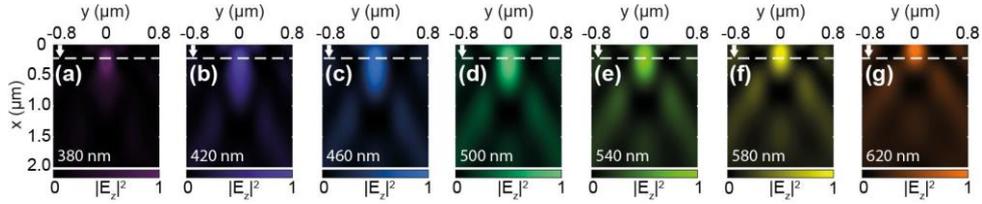

Fig. 7. The numerically calculated electric field intensity distributions with 40 nm wavelength steps at (a) 380 nm, (b) 420 nm, (c) 460 nm, (d) 500 nm, (e) 540 nm, (f) 580 nm, and (g) 620 nm. The dashed lines are representing average focal distance (=233 nm).

The electric field intensity ($|E_z|^2$) profiles of the focused beam in the *xy*-planes for selected operating wavelengths starting from 380 *nm* and ending at 620 *nm* (with 40 *nm* wavelength steps) are given in Fig. 7(a) - 7(g). These fields also verify the achromatic focusing effect of the proposed AMDL. Here, numerically calculated back focal distances are deviating from 100 *nm* to 280 *nm* for investigated visible wavelengths. From the given results, we can see that implemented optimization method accomplished the efficient focusing of light as well as achromatic focusing with the average focal length of 233 *nm* for the wavelengths between 380nm and 620nm.

## 5. Conclusion

In this study, we demonstrated an all-dielectric achromatic MDL design which is operating between 10 GHz and 14 GHz. The structural parameters of each concentric zones of the AMDL are optimized through 3D FDTD method combined with multi-objective differential evolution (MO-DE) algorithm. The optimized AMDL was fabricated via 3D-printing by using low refractive index PLA ($\varepsilon_{PLA} = 2.4025$) material to conduct microwave experiments for verification of the numerically obtained results. The fractional change between the average focal distances and the optimized desired focal length is measured 2.2% and 7.01% numerically and experimentally, respectively. Besides, the focusing efficiency values are changing from 28% to 45% in numerical results. To numerically demonstrate the performance of the AMDL, simulations are performed between the wavelengths between 380 nm – 620 nm by scaling down the optimized AMDL. The designed AMDL also shows near unity NA values over a broadband range. We believe that a proper design of this AMDL for visible and near infra-red applications may provide excellent solutions for focusing issues in terms of efficient and aberration-free focusing.

**Funding.** TED University Scientific Research Project with Project No: T-19-B2010-90016.

**Acknowledgments.** This work supported by H.K. acknowledges partial support from the Turkish Academy of Sciences.

**Disclosures.** The authors declare no conflicts of interest.

**Data availability.** No data were generated or analyzed in the presented research.